\documentclass[twoside,slac_one]{revtex4}
\usepackage{graphicx}
\usepackage{fancyhdr}
\usepackage{amsmath} 
\usepackage{bm}
\usepackage{amsxtra}
\usepackage{amssymb}
\usepackage{amsthm}
\usepackage{latexsym}
\usepackage{lscape}

\begin{document}

\title{Search for CP violation in the $B_{s} - \overline{B}_{s}$ system with LHCb}

%

\author{D. van Eijk on behalf of the LHCb collaboration}
\affiliation{Nikhef, Science Park 105, 1098 XG Amsterdam, The Netherlands}

\begin{abstract}
LHCb is one of the four large experiments at the Large Hadron Collider (LHC) and dedicated to heavy flavour physics. LHCb searches for New Physics (NP) by performing precision measurements of CP-violating parameters. One of the key measurements of LHCb is the extraction of the CP-violating weak phase $\phi_{s}$. The Standard Model (SM) prediction for this parameter is small: $\phi_{s}^{SM} = -0.0363 \pm 0.0017$ \cite{CKMFitter}. Possible deviations from this SM prediction may be attributed to NP.

$\phi_{s}$ is measured by doing a time-dependent angular analysis in the $B_{s}^{0} \rightarrow J/\psi\,\varphi$ channel. Key results on the road to such a measurement of $\phi_s$ are presented here. 

First, effective lifetimes in various $b \rightarrow J/\psi\,X$ channels are presented. Subsequently, decay amplitudes in $P \rightarrow VV$ decays are extracted from a time-dependent angular analysis in $B_{d}^{0} \rightarrow J/\psi\,K^{*}$ and $B_{s}^{0} \rightarrow J/\psi\,\varphi$ decays. In addition, the latter decay gives access to $\Delta \Gamma_{s}$, which is measured to be $\Delta \Gamma_{s} = 0.084\,\pm\,0.112\,(\mathrm{stat.})\,\pm\,0.021\,(\mathrm{syst.})\,\mathrm{ps}^{-1}$, assuming $\phi_{s}=0$.

A crucial ingredient for a $\phi_{s}$ measurement is the determination of the flavour of the $B_{s}^{0}$ meson at the time of production, a procedure called flavour tagging. Tagging information is also used in the measurement of the mixing frequency in the $B_{d}^{0}$ and $B_{s}^{0}$ system: $\Delta m_{d} = 0.499\,\pm\,0.032\,(\mathrm{stat.})\,\pm\,0.003\,(\mathrm{syst.})\,\mathrm{ps}^{-1}$ and $\Delta m_{s} = 17.63\,\pm\,0.11\,(\mathrm{stat.})\,\pm\,0.04\,(\mathrm{syst.})\,\mathrm{ps}^{-1}$ respectively.

In addition, the branching ratio of $B_{s}^{0} \rightarrow J/\psi f_{0}(980)$ is measured, which is another interesting channel to measure $\phi_{s}$.

Finally, results on two penguin decays are shown: first observation of $B_{s}^{0} \rightarrow K^{*}\overline{K^{*}}$ and evidence for $B_{s}^{0} \rightarrow J/\psi \overline{K^{*}}$.

All results are extracted from the 2010 LHCb data set which has an integrated luminosity of 37 pb$^{-1}$.
\end{abstract}

\maketitle

\thispagestyle{fancy}


\section{Introduction}
\begin{figure}[ht]
\centering
\begin{picture}(400,110)(0,0)
\put(20,30){\includegraphics[width=45mm]{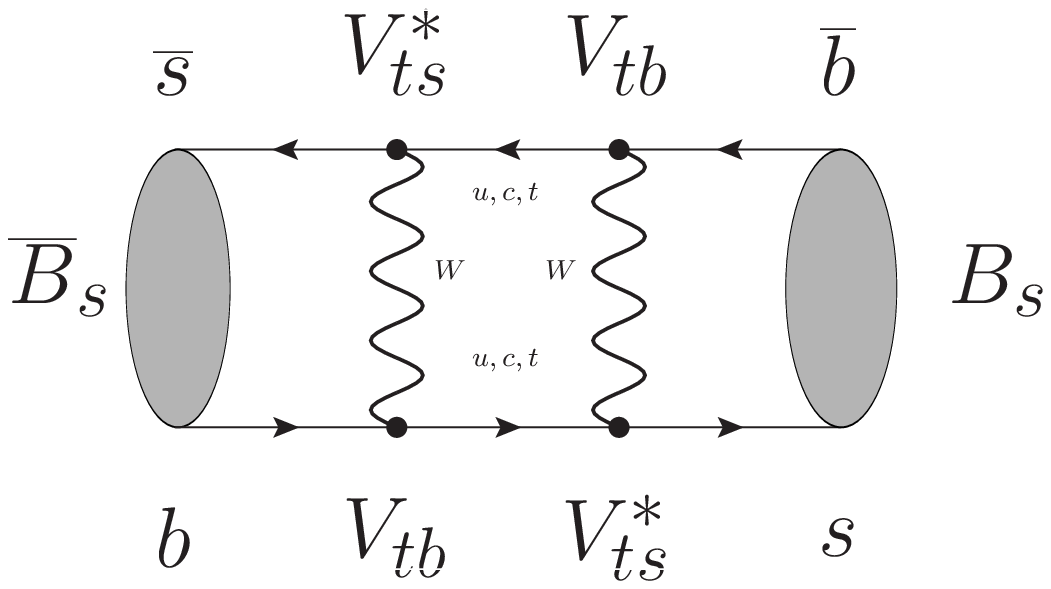}}
\put(20,10){ (a) $B_{s}^{0}$ mixing diagram}
\put(220,10){\includegraphics[width=60mm]{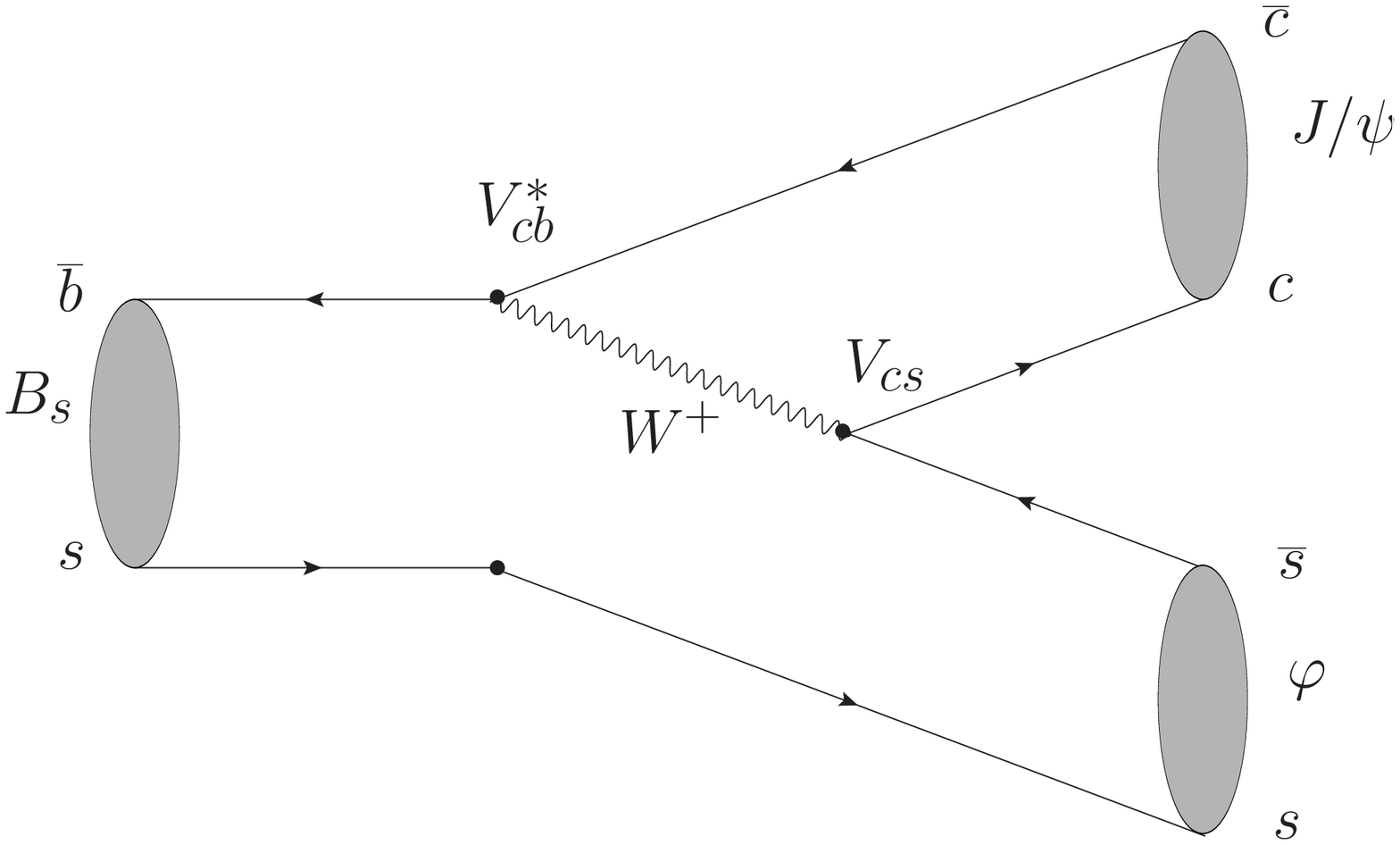}}
\put(240,10){ (b) Tree level decay}
\end{picture}
\caption{The $B_{s}^{0} \rightarrow J/\psi\,\varphi$ decay can occur via a direct decay (b) or via mixing (a) followed by the direct decay.} \label{fig:mixing}
\end{figure}

Transitions between the meson flavour eigenstates $B_{s}^{0}$ and $\overline{B_{s}^{0}}$ is a process known as mixing. The leading order diagram for $B_{s}^{0}$ mixing is shown in Fig. \ref{fig:mixing}(a). As a result, the flavour eigenstates differ from the mass eigenstates $B_{L}$ and $B_{H}$, which have masses $m_{L}$ and $m_{H}$, respectively. Their mass difference determines the $B_{s}^{0}$ mixing frequency $\Delta m_{s} = m_{H}-m_{L}$. The mass eigenstates have different total decay widths $\Gamma_{L}$ and $\Gamma_{H}$. In the SM the width difference $\Delta \Gamma_{s}=\Gamma_{L}-\Gamma_{H}$ is expected to be sizeable: $\Delta \Gamma_{s} = 0.087\,\pm 0.021\,\mathrm{ps}^{-1}$ \cite{deltagammaSM}.

The CP-violating phase $\phi_{s}$ is related to the interference between direct decay amplitudes and amplitudes of decay after mixing (see Fig. \ref{fig:mixing}). When neglecting penguin contributions, which are expected to be small \cite{fleischersmallpenguins}, the SM prediction for this phase is $\phi_{s}^{SM} = -2\beta_{s}$, where $\beta_{s} = \arg{(-V_{ts}V_{tb}^{*}/V_{cs}V_{cb}^{*})}$ is one of the angles in the $b-s$ unitarity triangle. New Physics contributions in the box diagram could enhance this phase: $\phi_{s} = \phi_{s}^{SM} \rightarrow \phi_{s} = \phi_{s}^{SM}+\Delta \phi_{s}^{NP}$. This means that measuring $\phi_{s}$ to high precision is a good way to search for NP.

The Tevatron has measured the $B_{s}^{0}$ mixing frequency $\Delta m_{s} = 17.77\,\pm\,0.10\,(\mathrm{stat.})\,\pm\,0.07\,(\mathrm{syst.})\, \mathrm{ps}^{-1}$ with high precision \cite{tevatrondeltams}, however the limits on $\Delta \Gamma_{s}$ and $\phi_{s}$ are less stringent \cite{tevatronphis}. 

LHCb is a dedicated heavy flavour experiment at the LHC, and one of its main goals is improving these measurements using the large number of produced $B$ mesons in the LHC collisions. 

\section{Towards a $\phi_{s}$ measurement}
To measure the CP asymmetry that is sensitive to $\phi_s$ one needs to have information on the flavour of the produced $B_{s}^{0}$ meson (flavour tagging) and on the CP eigenvalue of the final state.

The CP asymmetry is diluted by detector effects such as proper time resolution and imperfect flavour tagging. It is crucial to understand and quantify these detector effects. The $\phi_s$ measurement therefore consists of several steps:

\begin{itemize}
\item Define a common selection for $b \rightarrow J/\psi \, X$ channels and measure their effective lifetimes. In this analysis also the proper time resolution function is determined.
\item Perform an angular analysis in $B_{d}^{0} \rightarrow J/\psi K^{*}$ to disentangle the different CP eigenstates and measure $P \rightarrow VV$ transversity amplitudes and phases.
\item Perform an untagged analysis of $B_{s}^{0} \rightarrow J/\psi\,\varphi$ assuming $\phi_{s} = 0$ to measure $\Delta \Gamma_{s}$.
\item Calibrate the flavour tagging algorithm and use tagging information to measure $\Delta m_{d}$ and $\Delta m_{s}$ in $B\rightarrow D\pi$ and $B_{s}\rightarrow D_{s}\pi$ channels respectively.
\item Finally, measure $\phi_{s}$ in an unbinned time-dependent angular analysis of tagged $B_{s}^{0} \rightarrow J/\psi\,\varphi$ events.
\end{itemize}
In the next sections, these steps towards a measurement of $\phi_{s}$ are discussed in more detail.

\section{$b \rightarrow J/\psi \, X$ lifetime measurements}
$B$-hadron lifetimes are measured in the decays $B^{+} \rightarrow J/\psi K^{+}$, $B_{d}^{0} \rightarrow J/\psi K^{*}$, $B_{d}^{0} \rightarrow J/\psi K_{S}$, $B_{s}^{0} \rightarrow J/\psi\,\varphi$ and $\Lambda_{b} \rightarrow J/\psi\Lambda$. Signal yields and lifetimes resulting from a single exponential fit to the reconstructed proper time distributions are summarized in Table \ref{tab:lifetimes}. The result for $B_{s}^{0}$ is an effective lifetime, since the proper time distribution consists of two components: one for $B_{L}$ and one for $B_{H}$. The results are compatible with world averages, but don't have competitive errors yet. However these lifetime studies provide valuable input for the $\phi_{s}$ measurement, such as proper time resolution and acceptance.

The proper time resolution is modelled as the sum of three Gaussians with common mean but different width. The dilution of the CP asymmetry as a result of proper time resolution is $\mathcal{D} = \exp{(-\Delta m_{s}^{2} \sigma_{t}^{2}/2)}$ \cite{Moser}. For $B_{s}^{0} \rightarrow J/\psi\,\varphi$, the dilution found from the three Gaussians corresponds to an effective proper time resolution of $<\sigma_{t}>=50\,\mathrm{fs}^{-1}$. 

In the 2010 data a small linear drop in the efficiency is seen for large proper times. The exact shape of this acceptance curve is extracted from MC. More details can be found in \cite{lhcblifetimes}.

\begin{table}[t]
\begin{center}
\caption{Measured $b \rightarrow J/\psi \, X$ lifetimes \cite{lhcblifetimes}.}
\begin{tabular}{|l|l|c|l|}
\hline 
Decay channel & Yield & LHCb result $\tau$\,[ps] (value $\pm$ stat. $\pm$ syst.) & PDG $\tau$\,[ps] \\
\hline
$B^{+} \rightarrow J/\psi K^{+}$ & 6741 $\pm$ 85 & $1.689 \pm 0.022 \pm 0.047$ & $1.638 \pm 0.011$ \\ 
$B_{d}^{0} \rightarrow J/\psi K^{*}$ & 2668 $\pm$ 58 & $1.512 \pm 0.032 \pm 0.042$ & $1.525 \pm 0.009$ \\
$B_{d}^{0} \rightarrow J/\psi K_{S}$ & 838 $\pm$ 31 & $1.558 \pm 0.056 \pm 0.022$ & $1.525 \pm 0.009$ \\
$B_{s}^{0} \rightarrow J/\psi\,\varphi$ & 570 $\pm$ 24 & $1.447 \pm 0.064 \pm 0.056$ & $1.472^{+0.024}_{-0.026}$ \\
$\Lambda_{b} \rightarrow J/\psi\Lambda$ & 187 $\pm$ 16 & $1.353 \pm 0.108 \pm 0.035$ & $1.391^{+0.038}_{-0.037}$ \\
\hline
\end{tabular}
\label{tab:lifetimes}
\end{center}
\end{table}

\section{Angular analysis of the  $B_{d}^{0} \rightarrow J/\psi K^{*}$ decay}

$B_{d}^{0} \rightarrow J/\psi K^{*}$ is a decay of a pseudo-scalar to two vector mesons ($P \rightarrow VV$). This type of decays can be described in terms of three decay amplitudes: $A_{0}$, $A_{\perp}$ and $A_{\parallel}$. These amplitudes are related to the possible relative angular momentum configurations of the vector mesons in the final state: $L=0,1,2$ respectively. $\delta_{\parallel}$ and $\delta_{\perp}$ denote the strong phases of $A_{\perp}$ and $A_{\parallel}$ relative to $A_{0}$. The extracted amplitudes and phases are given in Table \ref{tab:Bdangularanalysis} and agree within their errors with measurements done by other experiments.

This result is an important cross-check for the correct implementation of angular dependent acceptance effects caused by detector geometry and event selection. MC data is used to assess this angular acceptance effect.

The possible presence of an S-wave contribution was accounted for in the fit. Systematic uncertainties that have been evaluated include the background shape, acceptance effects and the signal mass model. More details can be found in \cite{untaggedanalysis}.

\begin{table}[t]
\begin{center}
\caption{Results from an untagged angular fit to $B_{d}^{0} \rightarrow J/\psi K^{*}$ events \cite{untaggedanalysis}.}
\begin{tabular}{|c|c|}
\hline 
Parameter & LHCb result (value $\pm$ stat. $\pm$ syst.)\\
\hline
$|A_{\parallel}|^2$ & $0.252 \pm 0.020 \pm 0.016$ \\
$|A_{\perp}|^2$ & $0.178 \pm 0.022 \pm 0.017$ \\
$\delta_{\parallel}$ & $-2.87 \pm 0.11 \pm 0.10$ \\ 
$\delta_{\perp}$ & $3.02 \pm 0.10 \pm 0.07$ \\ 
\hline
\end{tabular}
\label{tab:Bdangularanalysis}
\end{center}
\end{table}

\section{Untagged analysis of the $B_{s}^{0} \rightarrow J/\psi\, \varphi$ decay}
The decay $B_{s}^{0} \rightarrow J/\psi\, \varphi$ is a $P \rightarrow VV$ decay as well and thus is very similar to $B_{d}^{0} \rightarrow J/\psi K^{*}$ which was discussed in the previous section. $\Delta \Gamma_{s}$ is the decay width difference between the heavy and light $B_{s}$ mass eigenstates and can be measured in an untagged angular analysis of $B_{s}^{0} \rightarrow J/\psi\, \varphi$ events. This means no information on the flavour of the initial $B_{s}$ meson is used.

In the untagged analysis $\phi_{s}=0$ is assumed which is close to the Standard Model prediction. The results are given in Table \ref{tab:Bsangularanalysis}. 

\begin{figure}[ht]
\centering
\includegraphics[width=40mm, angle=270]{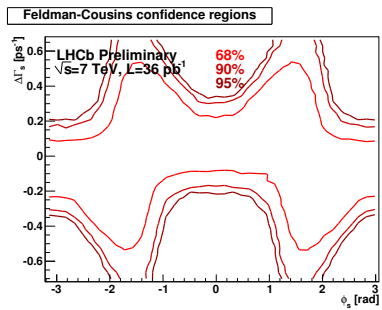}
\caption{$\phi_{s}-\Delta \Gamma_{s}$ confidence contours of the untagged angular analysis of $B_{s}^{0} \rightarrow J/\psi \,\varphi$ events. The contours are determined by performing a Feldman-Cousins study with 1000 toys for each grid point (40 $\times$ 40 grid points). $\phi_{s}$ can clearly not be constrained by doing an untagged analysis with the statistics of the 2010 data \cite{untaggedanalysis}.} \label{fig:feldmancousinsuntagged}
\end{figure}

Systematic uncertainties that have been studied include the background shape, the angular acceptance and a possible S-wave contribution. To estimate the sensitivity of the untagged fit to $\phi_{s}$ a Feldman-Cousins study \cite{feldmancousins} is performed in the two dimensional $\phi_{s}-\Delta \Gamma_{s}$ plane. The resulting confidence contours are shown in Fig. \ref{fig:feldmancousinsuntagged}. This figure shows that with the current statistics $\phi_{s}$ can not be constrained when performing an untagged analysis. For the extraction of $\phi_{s}$ tagging information is needed. 

\begin{table}[t]
\begin{center}
\caption{Results from an untagged angular fit to $B_{s}^{0} \rightarrow J/\psi \,\varphi$ events. In this fit, $\phi_{s}=0$ is assumed \cite{untaggedanalysis}.}
\begin{tabular}{|c|c|}
\hline 
Parameter & LHCb result (value $\pm$ stat. $\pm$ syst.)\\
\hline
$\Gamma_{s}$ [ps$^{-1}$] & $0.680 \pm 0.034 \pm 0.027$ \\
$\Delta \Gamma_{s}$  [ps$^{-1}$] & $0.084 \pm 0.112 \pm 0.021$ \\
$|A_{0}|^2$ & $0.532 \pm 0.040 \pm 0.028$ \\ 
$|A_{\perp}|^2$ & $0.279 \pm 0.057 \pm 0.014$ \\ 
$\cos \delta_{\parallel}$ & $-1.24 \pm 0.27 \pm 0.09$ \\
\hline
\end{tabular}
\label{tab:Bsangularanalysis}
\end{center}
\end{table}

\section{Tagging}
\begin{figure}[ht]
\centering
\includegraphics[width=100mm, trim= 0 0 0 1, clip=true]{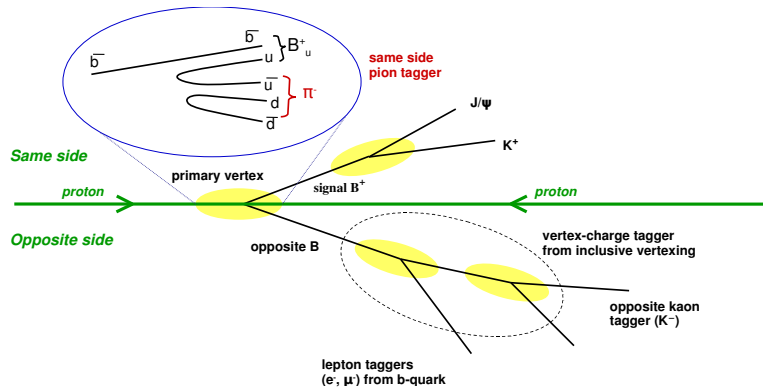}
\caption{Schematic picture of tagging in LHCb.} \label{fig:tagschema}
\end{figure}

There are two types of taggers in LHCb (see figure \ref{fig:tagschema}): opposite side (OS) and same side (SS) taggers. 

The OS tagger algorithms (muon, electron and kaon) use the charge of the lepton from semileptonic $b$-hadron decays and the kaon from the $b \rightarrow c \rightarrow s$ decay chain to define the flavour of the signal $B$ meson. The inclusive secondary vertex tagger uses the charge of the inclusive secondary vertex reconstructed from $b$-decay products.

The SS tagger algorithms directly determine the flavour of the signal $B$ meson. In $B_d^0 / B_s^0$ events, from the fragmentation of a $\overline{b}$ quark, an extra $d/s$ is available to form a pion/kaon that in a fraction of the cases is charged and identifies the flavour of the signal $B$.

In summary, OS taggers (muon, electron, kaon and inclusive secondary vertex) can be used to tag any $b$-hadron, whilst SS pion (SS$\pi$) and SS kaon (SS$K$) taggers can be used only to tag $B^{0}$ and $B^{+}$ or $B^0_s$ , respectively. For each tagger, the probability of the tag decision to be correct is estimated by means of a neural network trained on MC events to identify the correct flavour of the signal $B$ meson. These probabilities are then calibrated on data using the self-tagging $B^+ \rightarrow J/\psi K^+$ control channel.

The sensitivity of a CP asymmetry is directly related to the effective tagging power $\epsilon_{\mathrm{eff}} = \epsilon_{\mathrm{tag}} (1-2\omega)^2 = \epsilon_{\mathrm{tag}} D^2 $, where $\omega$ is the mistag probability as returned by the tagging algorithm and $\epsilon_{\mathrm{tag}}$ is the tag decision efficiency. This tagging power represents the effective statistical reduction of the data sample due to mistag probability. 

With the 2010 dataset, the SS tagger is not calibrated yet, so for the $\phi_{s}$ measurement only the OS tagger is used. In the $B_{s}^{0} \rightarrow J/\psi\, \varphi$ analysis the effective mistag probability is $\omega_{\mathrm{eff}} = 32\% \pm 2\%$, which results in a tagging power of $\epsilon D^2 = 2.2\% \pm 0.5\%$. 

\section{$\Delta m_{d,s}$ measurement}
The mixing frequency in the $B_{d}^{0}$ system is determined from the time-dependent mixing asymmetry in the decay $B_{d}^{0}\rightarrow D \pi$, defined as $A(t)=(N_{\mathrm{unmixed}}(t)-N_{\mathrm{mixed}}(t))/(N_{\mathrm{unmixed}}(t)+N_{\mathrm{mixed}}(t))$. The measured value is $\Delta m_{d} = 0.499\,\pm\,0.032\,(\mathrm{stat.})\,\pm\,0.003\,(\mathrm{syst.})\,\mathrm{ps}^{-1}$ which is in agreement with the world average $\Delta m_{d} = 0.507\,\pm\,0.005\,\mathrm{ps}^{-1}$. More details can be found in \cite{deltamdresult}.

\begin{figure}[ht]
\centering
\begin{picture}(400,130)(0,0)
\put(20,20){\includegraphics[width=55mm]{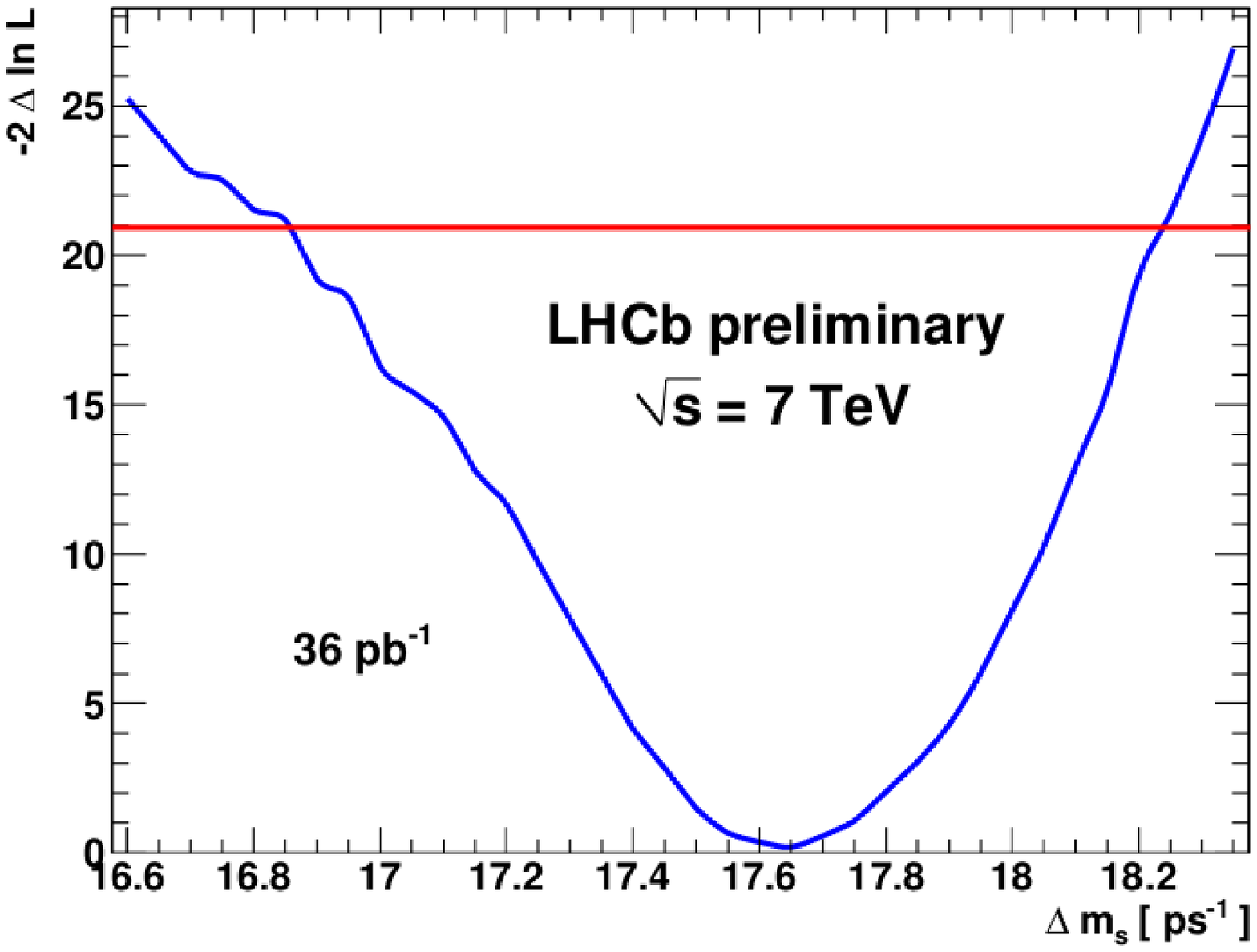}}
\put(10,10){ (a) Likelihood scan as function of $\Delta m_{s}$}
\put(220,20){\includegraphics[width=55mm]{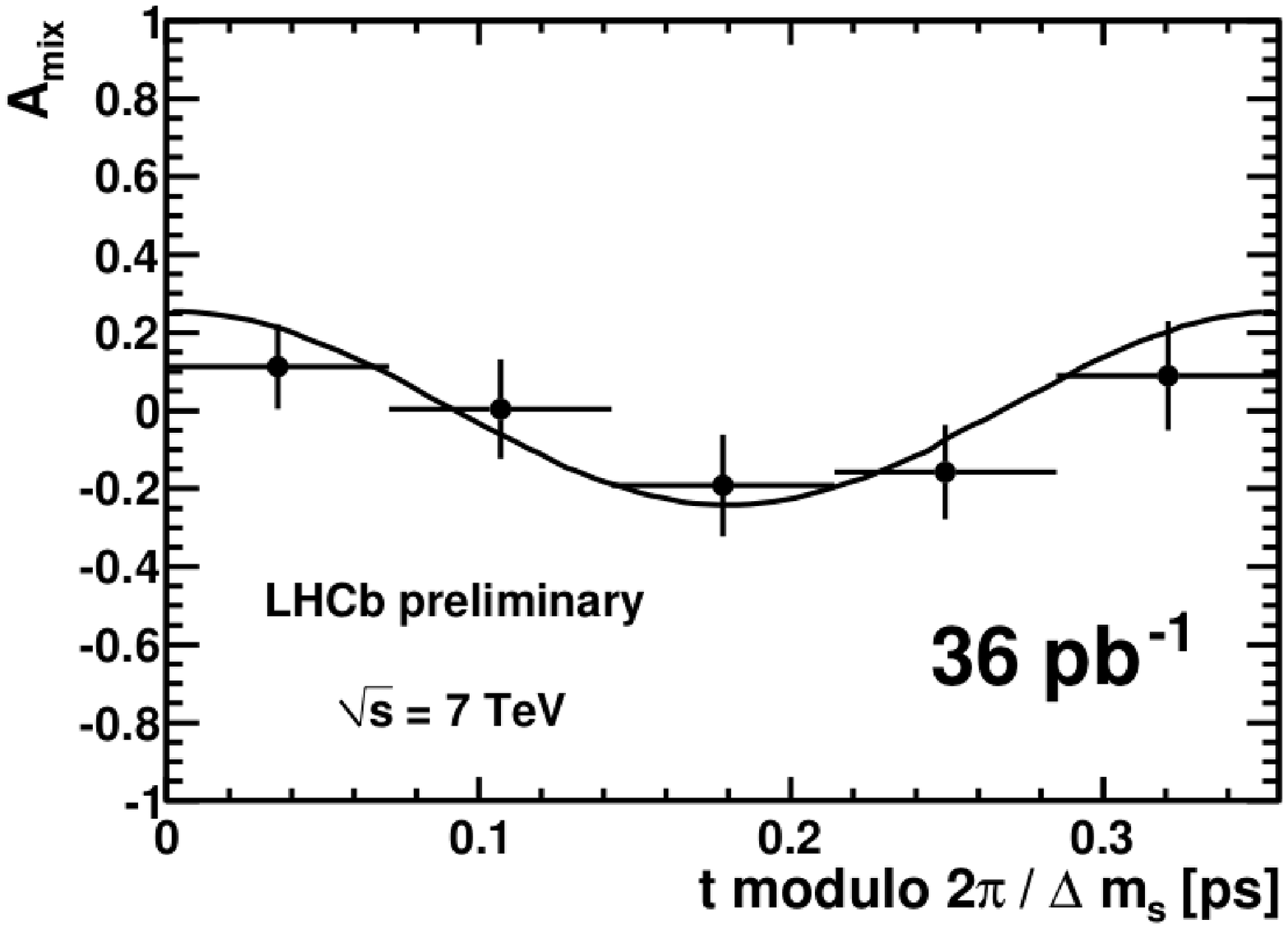}}
\put(250,10){ (b) $B_{s}^{0}$ mixing assymetry}
\end{picture}
\caption{(a) Likelihood scan as a function of  $\Delta m_{s}$ showing the 4.6$\sigma$ significant mixing in the $B_{s}^{0}$ system. (b) Time dependent mixing asymmetry modulo $2\pi/\Delta m_{s}$. Both plots are taken from \cite{deltamsresult}.} \label{fig:deltams}
\end{figure}

Using $B_{s}^{0} \rightarrow D_{s}\pi,D_{s}\pi\pi\pi$ decays, mixing in the $B_{s}^{0}$ system has been observed with a significance of 4.6$\sigma$ and the mixing frequency is measured as $\Delta m_{s} = 17.63\,\pm\,0.11\,(\mathrm{stat.})\,\pm\,0.04\,(\mathrm{syst.})\,\mathrm{ps}^{-1}$. This measurement is competitive with the measurement published by the CDF collaboration: $\Delta m_{s} = 17.77\,\pm\,0.10\,(\mathrm{stat.})\,\pm\,0.07\,(\mathrm{syst.})\, \mathrm{ps}^{-1}$ \cite{tevatrondeltams}. Since the $B_{s}^{0}$ oscillations are much faster than oscillations in the $B_{d}^{0}$ system, in this measurement LHCb profits from its excellent proper time resolution and is able to overcome the statistical limitation with the 2010 dataset. More details are given in \cite{deltamsresult}. Fig. \ref{fig:deltams} shows the likelihood scan as a function of $\Delta m_{s}$ and the time dependent mixing asymmetry modulo $2\pi/\Delta m_{s}$.

\section{Tagged analysis of the $B_{s}^{0} \rightarrow J/\psi\, \varphi$ decay}

\begin{figure}[ht]
\centering
\includegraphics[width=40mm,angle=270, trim= 0 0 0 1,clip=true]{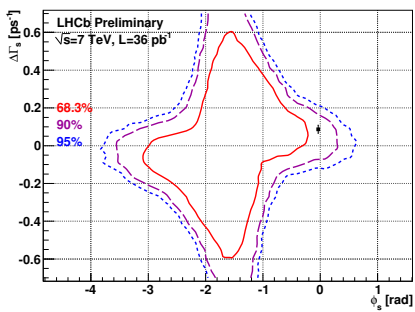}
\caption{Feldman-Cousins confidence contours in the $\phi_{s}-\Delta\Gamma_{s}$ plane. The CL at the SM point (black square) is 0.78 which corresponds to a p-value of 22\%. Projected in one dimension, one finds $\phi_{s} \in [-2.7, -0.5]\,\mathrm{ra}$d at 68\% CL \cite{phisresult}.} \label{fig:phis}
\end{figure}

Using the input of the steps described in the previous sections it is now possible to perform a tagged time-dependent angular analysis of $B_{s}^{0} \rightarrow J/\psi\, \varphi$ events. Unfortunately, using the 2010 dataset consisting of $836\pm60$ $B_{s}^{0} \rightarrow J/\psi\, \varphi$ events, no meaningful point estimates can be given for $\phi_s$ yet. Instead a Feldman-Cousins study has been performed in the $\phi_{s}-\Delta\Gamma_{s}$ plane, just as in the untagged fit. The results are shown in Fig. \ref{fig:phis}.

The probability of a fluctuation from the Standard Model expectation to the observed result for $\phi_{s}$ and $\Delta \Gamma_{s}$ is 22\%. A one-dimensional interval for $\phi_{s}$ is derived at 68\% CL: $\phi_{s} \in [-2.7, -0.5]\,\mathrm{rad}$. 

All studied systematic variations of the fitting conditions have an insignificant effect on the $\phi_{s}-\Delta \Gamma_{s}$ confidence contours. Therefore, the contours include only the statistical uncertainty, with the exception of the uncertainties due to flavour tagging calibration parameters and mixing frequency, which were floated in the fit. More details can be found in \cite{phisresult}.

\begin{figure}[ht]
\centering
\includegraphics[width=55mm]{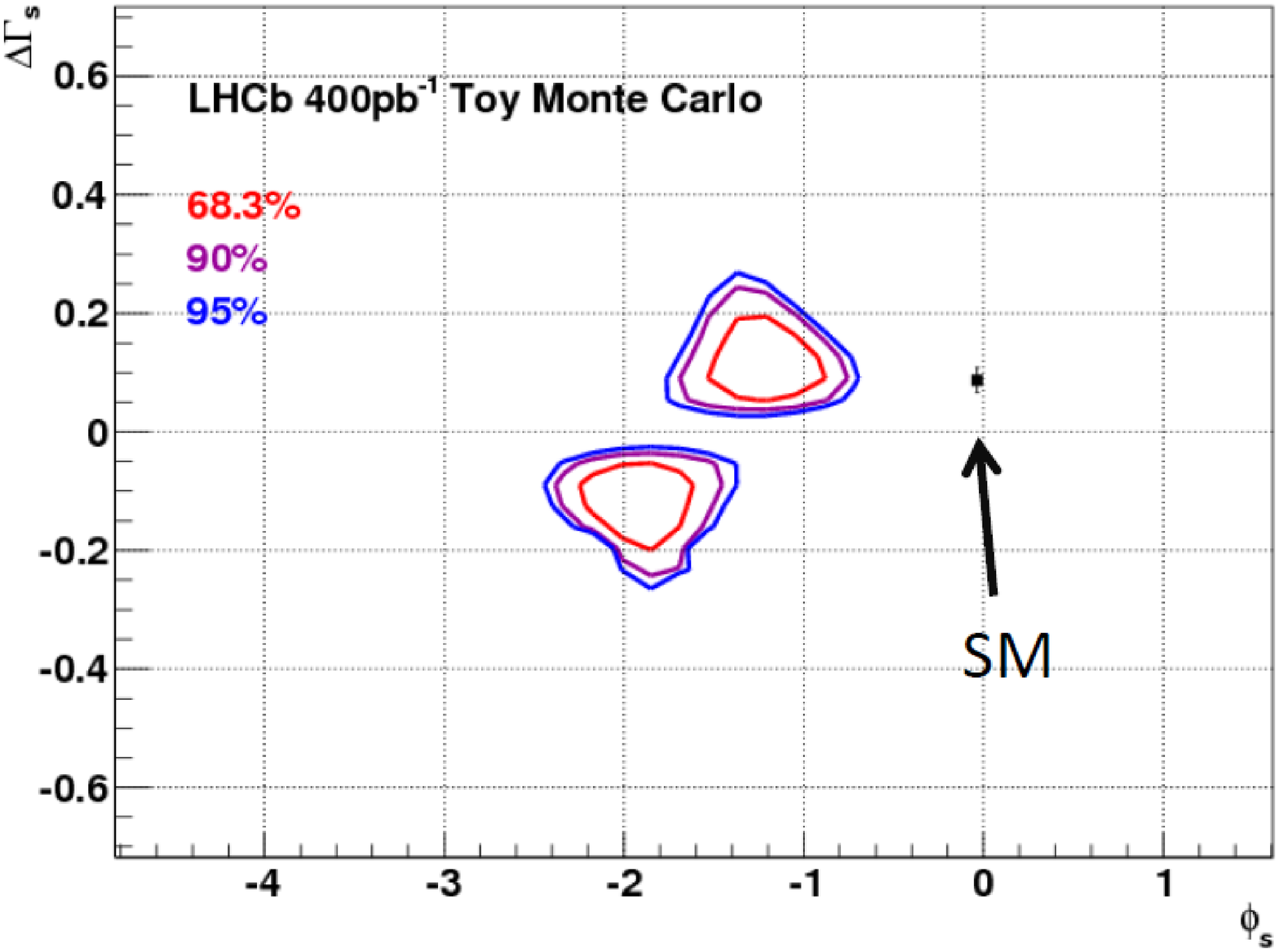}
\caption{Feldman-Cousins confidence contours in the $\phi_{s}-\Delta\Gamma_{s}$ plane for toy MC with an integrated luminosity of $400\,\mathrm{pb}^{-1}$. The conditions to generate were chosen the same as the values found in the 2010 analysis.} \label{fig:MCtoy}
\end{figure}

MC toys were generated using the same conditions as the 2010 fit but with an integrated luminosity of $400\,\mathrm{pb}^{-1}$ (expected for Summer 2011). The confidence contours are shown in Fig. \ref{fig:MCtoy}. Shortly after the DPF 2011 conference the LHCb result for $\phi_s$ was updated at the Lepton Photon 2011 conference, showing no significant deviation from the SM value.

\section{Other channels}
\begin{figure}[ht]
\centering
\includegraphics[width=55mm]{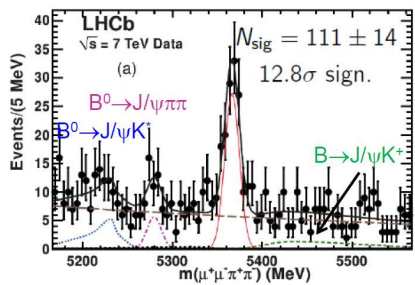}
\caption{$\mu^{+}\mu^{-}\pi^{+}\pi^{-}$ mass distribution showing the $B_{s}^{0} \rightarrow J/\psi f_{0}(980)$ mass peak \cite{jpsif0}.} \label{fig:jpsif0}
\end{figure}

In addition to the $\phi_s$ measurement in $B_{s}^{0} \rightarrow J/\psi\,\varphi$, LHCb also reports a first observation of $B_{s}^{0} \rightarrow J/\psi f_{0}(980)$ (see Fig. \ref{fig:jpsif0}), including a BR measurement \cite{jpsif0}: 
\begin{equation}
R_{f_{0}/\varphi} = \frac{\Gamma(B_{s}^{0}\rightarrow J/\psi f_{0}, f_{0}\rightarrow\pi^{+}\pi^{-}))}{\Gamma(B_{s}^{0}\rightarrow J/\psi\,\varphi, \varphi\rightarrow K^{+}K^{-}))} = 0.252^{+0.046+0.027}_{-0.032-0.033} \nonumber
\end{equation}
This channel can be used to measure $\phi_{s}$ as well. The advantage of this channel over $B_{s}^{0} \rightarrow J/\psi\,\varphi$ is that the final state has a definite CP eigenvalue. This means no angular analysis is needed to disentangle different CP eigenstates in the final state.

In addition, LHCb reports evidence for $B_{s}^{0} \rightarrow J/\psi \overline{K^{*}}$ with a branching ratio of $\mathcal{B}(B_{s}^{0} \rightarrow J/\psi \overline{K^{*}}) = (3.5^{+1.1}_{-1.0}(\mathrm{stat.})\pm0.9(\mathrm{syst.}))\times10^{-5}$ \cite{BsJpsiKstar}.

Finally, a first observation has been made of $B_{s}^{0} \rightarrow K^{*} \overline{K^{*}}$ with a branching fraction of $\mathcal{B}(B_{s}^{0} \rightarrow K^{*}\overline{K^{*}}) = (1.95 \pm 0.47(\mathrm{stat.}) \pm 0.51(\mathrm{syst.}) \pm 0.29(f_d/f_s))\times10^{-5}$ \cite{BsKstarKstar}.

\section{Summary}
The first steps on the way to a measurement of $\phi_{s}$ at LHCb have been presented. Using the data taken by the LHCb detector in 2010 ($37\,\mathrm{pb}^{-1}$ of integrated luminosity), polarization amplitudes and strong phases have been measured in $B_{d}^{0} \rightarrow J/\psi K^{*}$ decays. An untagged analysis of $B_{s}^{0} \rightarrow J/\psi\,\varphi$ events was performed to measure $\Delta \Gamma_{s}$. In addition the mixing frequencies $\Delta m_{d}$ and $\Delta m_{s}$ have been measured. 

Using the 2010 dataset no meaningful point estimates can be given for $\phi_{s}$ yet, so the result is presented as confidence contours in the $\phi_{s}-\Delta\Gamma_{s}$ plane. The updated LHCb $\phi_s$ result presented shortly after DPF 2011 at Lepton Photon 2011 shows no significant deviation from the SM value.

In addition branching fraction for three other interesting $B_{s}$ decays have been presented: $B_{s}^{0} \rightarrow J/\psi f_{0}(980)$, $B_{s}^{0} \rightarrow J/\psi \overline{K^{*}}$ and $B_{s}^{0} \rightarrow K^{*} \overline{K^{*}}$.

\begin{acknowledgments}
This document is partly based on proceedings for the Rencontres de Moriond: QCD and High Energy Interactions 2011 conference \cite{MoriondProceedings}.
\end{acknowledgments}

\bigskip 

\end{document}